# Wind, Sand and Water. The Orientation of the Late Roman Forts in the Kharga Oasis (Egyptian Western Desert)

**Corinna Rossi**
Politecnico di Milano, Department ABC
Campus Leonardo, Building 15
Via Ponzio 31, 20133 Milan, Italy
Tel. +39 02 2399 5157
Email: corinna.rossi@polimi.it

**Giulio Magli**
Politecnico di Milano, Department of Mathematics
Campus Leonardo, Building Nave
Piazza Leonardo da Vinci 32, 20133 Milan, Italy
Tel. +39 02 2399 4597
Email: giulio.magli@polimi.it

**Abstract**

The chain of late Roman fortified settlements built in the Kharga Oasis, in Egypt's Western Desert, represents an interesting case-study to analyse how the ancient Roman town planners interacted with the landscape. A peculiar feature of the site is the existence of a prevailing, north-westerly wind, and it is possible to identify the average azimuth of the wind by measuring the central axes of the half-moon shaped sand dunes which characterize the landscape. Using the methods of Archaeoastronomy, we compared these azimuths with the orthogonal layout of both the settlements and the agricultural installations and showed that these are oriented on the prevailing wind. A description and the possible implications of this 'weathervane orientation' are discussed in this article.

**Introduction**

The northern outskirts of Kharga, one of the five major oases of Egypt's Western Desert, are punctuated by a scatter of Late Roman installations that survive in relatively good conditions, thanks to the dry desert environment and their remote position. This network of sites, possibly belonging to the re-organisation of the empire's southern frontier triggered by Diocletian, was built along the most important caravan routes that met in the Kharga Oasis, a desert hub for trans-saharan commercial movements. They appear to have played a combination of roles: beside controlling the routes, they exploited the agricultural and mining potential of the peripheral areas of the oases, and demonstrated the presence and power of Rome thanks to their imposing and aggressive appearance (Reddé 1991; Tallet et al. 2012a; Rossi 2013;).

All these sites were endowed with large-scale agricultural systems, meant to make them self-sufficient (Tallet et al. 2011; Bravard et al. 2016; Rossi 2016; Rossi et al. forthcoming;). The sites located on lower ground, closer to the centre of the oasis' depression, were served by wells, whereas the ones located on higher ground mainly relied on subterranean aqueducts of the type called *qanawat* or *manawir*. The subterranean aqueducts of Umm al-Dabadib were the first archaeological feature of the entire area to be described in some detail by the British geologist and explorer H. Ll. Beadnell, who worked with J. Ball at the Geological Survey of Egypt at the very beginning of the 20th century (Ball 1900; Beadnell 1909).

In the majority of the Late Roman archaeological sites of north Kharga, the remains of the ancient fields still survive: the *centuriatio* can be clearly discerned from aerial and satellite images, and at ground level at dawn and sunset when the sun is very low (Rossi 2016). The buildings belonging to this chain of installations share a significant number of architectural features, suggesting that the builders drew from a common set of models (Rossi and Ikram forthcoming). We will focus here on the problem of the orientation of these archaeological sites.

It is indeed well known that, according to a variety of ancient sources, Roman city and land planning involved procedures inherited from the Etruscans and closely connected with the celestial cycles. For instance, the town's foundation ritual is described by Roman historians (such as Varro, Plutarch and Pliny the Elder) as a rule directly inherited from the Etruscans' sacred books of the aruspexes, the *Disciplina*. A fundamental part of the aruspexes' duty was connected with the cosmic order, and, as a consequence, a role for astronomy is to be expected in the orientation of the Roman layouts. In the literature, a number of cases have been actually documented (Magli 2006 and 2008; González-García et al. 2014; Bertarione and Magli 2015) and, in particular, cardinal orientation was one of the possible choices.

In spite of this, attempts at establishing a common rule must comply with the practical mentality of the Romans, which in many cases overruled symbolic principles in favour of local (e.g. geo-morphological) considerations. For instance, centuriations clearly inspired by astronomical considerations exist (for instance that of Roman Carthage), but other clearly inspired exclusively by the morphology of the terrain have been also documented. One example is Luni, in Italy, in the Massa Carrara province. The land division there was to be planned in a territory which is encased between the (roughly parallel) natural borders of the Ligurian sea and of the Apuan mountains. As a consequence, a grid not parallel to the shoreline would rapidly lead to many incomplete lots at both ends. The Roman planners thus opted for a purely practical solution and traced the main axis of the centuriation along the natural constraint line. In the present paper, we shall show that also the planning of the Roman settlements in Kharga is very likely an example of ori-entation purely based on a practical solution: in this case, the need to take into account the direction of the prevailing wind.

**The Late Roman Settlements in northern Kharga**

In comparison with the Nile Valley, the archaeological remains of the oases received little attention until the Egyptian Egyptologist Ahmed Fakhry embarked in a tour of the area in the '40s. His reports were published in the subsequent decades (Fakhry 1950, 1951, 1973; his notes on the Kharga Oasis remain unpublished). A more systematic exploration arrived later, together with the first asphalted road, that reached Kharga in the '60s. The archaeological sites located closer to the inhabited parts of the oasis were studied first (Fakhri 1951; Cruz-Uribe 1986 and 1987; Reddé at al. 2004). They were followed by the areas located at the outskirts (e.g. Gascou et al. 1979; Dunand et al. 2003), by the most isolated sites (Rossi 2000; Rossi and Ikram 2010) and eventually by all the elusive remains such as the agricultural installations associated to these sites (Wuttmann et al. 2000; Newton et al. 2013; Rossi 2016). All these studies are slowly building a coherent picture of the situation of the Kharga Oasis in the Roman period (Wagner 1987, Reddé 1991, Tallet et al. 2012b; Rossi and Ikram forthcoming).

Further work will be necessary in the future to address the earlier historical periods, still poorly represented. The history of the Kharga Oasis during the pharaonic period is, in fact, still obscure. Prehistory left abundant traces on the now-desert outskirts of the depression, which was once occupied by a large lake (Caton-Thompson 1952). The pharaonic remains, instead, are likely to have been concentrated in the lowest area of the depression and to have been buried by the continuous occupation of the greenest portion of the oasis, throughout the subsequent millennia. Only a few remains dating to this long period have been so far retrieved from the core of the oasis (Cappers at al. 2014), but the caravan routes that crossed the region offer substantial evidence on the existence of a network of transport and communications, suggesting a continuous and significant occupation of the area through all that historical period (Rossi and Ikram 2015; Rossi and Ikram 2013; Ikram 2006; Ikram and Rossi 2004a; Rossi and Ikram 2002).

Late Period remains are limited, but significant: the earliest core of the Hibis Temple dates to this period, and indicates a strong interest for this area (Cruz-Uribe 1987 and 1986; Winlock 1941). The scale of the Ptolemaic exploitation of the Western Desert oases is unclear, but might have been more substantial than previously thought (Gill 2016).

The oasis is clearly punctuated by well-preserved archaeological remains indicating that the Romans poured significant resources into the construction of settlements; in particular, a chain of military-looking installations was built in the Late Roman period. In northern Kharga they are, from north to south: the little forts of Qasr al-Gib and Qasr al-Sumayra (Ikram and Rossi 2004b), the fort of Mohammed Tuleib, the legionary fortress of al-Deir (Brones and Duvette 2008), and two similar, rather substantial installations: the Fort and the surrounding Gridded Settlement at Ain al-Lebekha (Rossi and Ikram 2010), and the Fort and the Fortified Settlement at Umm al-Dabadib (Rossi and Ikram 2006; Rossi 2000).

Near the central part of the oasis' depression, two more sites might belong to this chain: Qasr al-Nessima, and Qasr al-Baramoudy. They have never been studied and lie in an area that is not currently covered by good resolution satellite images. In general, they strongly resemble Umm al-Dabadib and Qasr al-Lebekha in terms of general layout, as they both consists of a central sturdy building surrounded by a dense settlement laid out on a gridded pattern.

To the Roman period, broadly speaking, also belong three sites consisting of mudbrick enclosures surrounding sizeable stone temples: Nadura, Qasr al-Ghweita and Qasr al-Zayyan, that have been partly investigated (Klotz 2013; Cappers et al. 2014; Klotz 2009). In the south of the oasis, the most important Roman installation is the large site of Dush. This vast and stratified site had earlier origins: the nearby site of Ain al-Manawir appears to have been occupied from the Persian period onwards, and the $4^{th}$ century AD Roman military base of Dush is likely to have re-used the structure of older fortified magazines (Reddé et al. 2004).

We shall now focus on the orientation of the Late Roman settlements belonging to the chain that was built in the northern and central portion of Kharga.

**Orientation(s)**

In Kharga, only small portions of the extant archaeological complexes have been excavated and surveyed in a detailed way. This is due to a combination of reasons: first of all, because large-scale excavation can be extremely complicated in such a sandy environment (e.g. Reddé at al. 1990: 284); then because the complex logistic conditions dictate the adoption of some methods instead of others (e.g. Fassi et al. 2015); and finally because of the lack of fixed points to georeference the surveys in the most remote sites. This means that we do not possess a large number of precise surveys, and that the orientation of several settlements and installations has been so far measured in an approximate way. The Hibis Temple represents an exception, as it has been the object of a careful conservation project.

We shall now focus on the chain of Late Roman fortified settlements that punctuate the Kharga Oasis, namely Qasr al-Gib, Qasr al-Sumayra, Muhammad Tuleib, Qasr al-Lebekha, Umm al-Dabadib, al-Deir, Qasr al-Nessima, and Qasr al-Baramoudy. The general lack of precise measurements can be circumvented by using Google Earth. The current resolution of the available images allows a measurement of the orientation of the visible buildings with a tolerance of about one degree, which is by far adequate for the current discussion (figure 1). All the structures are square or rectangular in plan, and therefore it is sufficient to measure the azimuth of one of their sides. The results are reported in Table 1.

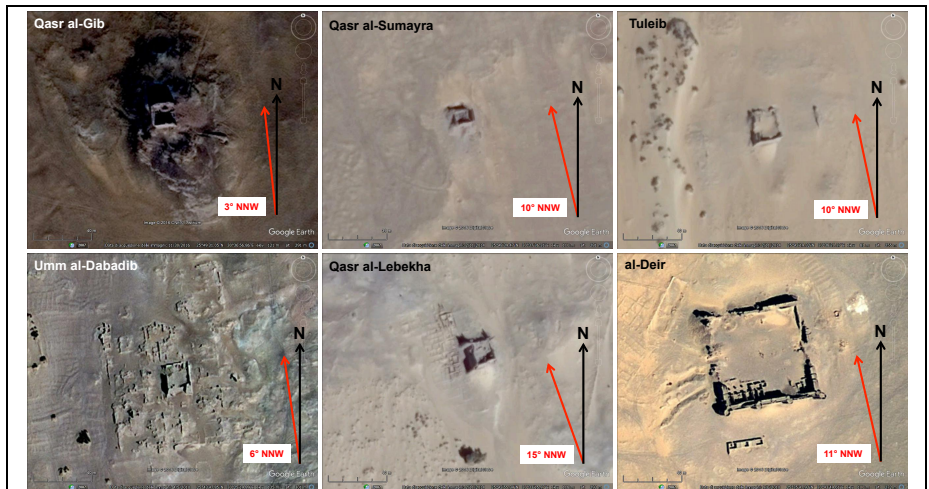

**Figure 1**: Google Earth images of six Late Roman settlements in northern Kharga and their orientation.

| Name | Azimuth of built-up areas | Approximate azimuth of barchan dunes |
|---|---|---|
| Qasr al-Gib | 3° W | ~10°W |
| Qasr al-Sumayra | 10° W | ~10°W |
| Tuleib | 10° W | ~10°W |
| Umm al-Dabadib | 6° W | ~7°W |
| Qasr al-Lebekha | 15° W | ~14°W |
| Al-Deir | 11° W | ~11°W |
| Qasr al-Baramoudy | 10° E | ~10°E |
| Qasr al-Nessima | 7° E | ~7°E |
| Dush | 15° W | 8-15 W |
| Dush Settlement | 8°W | 8-15 W |

Table 1: orientation of the Late Roman fortified sites of Kharga.

The data cluster around true north, but we can exclude cardinal orientation as in this case the errors committed by the builders would have been too high. A first key to interpret these data is offered by the Fortified Settlement of Umm al-Dabadib and its associated cultivations (figure 2): the satellite images clearly show that the orientation of the Fortified Settlement and of the *centuriatio* is the same as the average orientation of the sand dunes. Dunes are shaped by the wind, and *barchan* (or half-moon shaped) dunes, as the ones at Dabadib, develop when the wind always blows from the same direction; it is therefore meaningful to speak of average orientation of these natural features by taking the (of course approximate) azimuth of the central axis of the arch-shaped form. These dunes are persistent features of the landscape: the size of the area covered by them and the relatively slow speed at which they move indicate that the wind has been blowing in the same way for several thousands of years.

The Fortified Settlement and its associated cultivations were therefore oriented *secundum naturam*: the natural feature to be taken into account here being the prevailing wind.

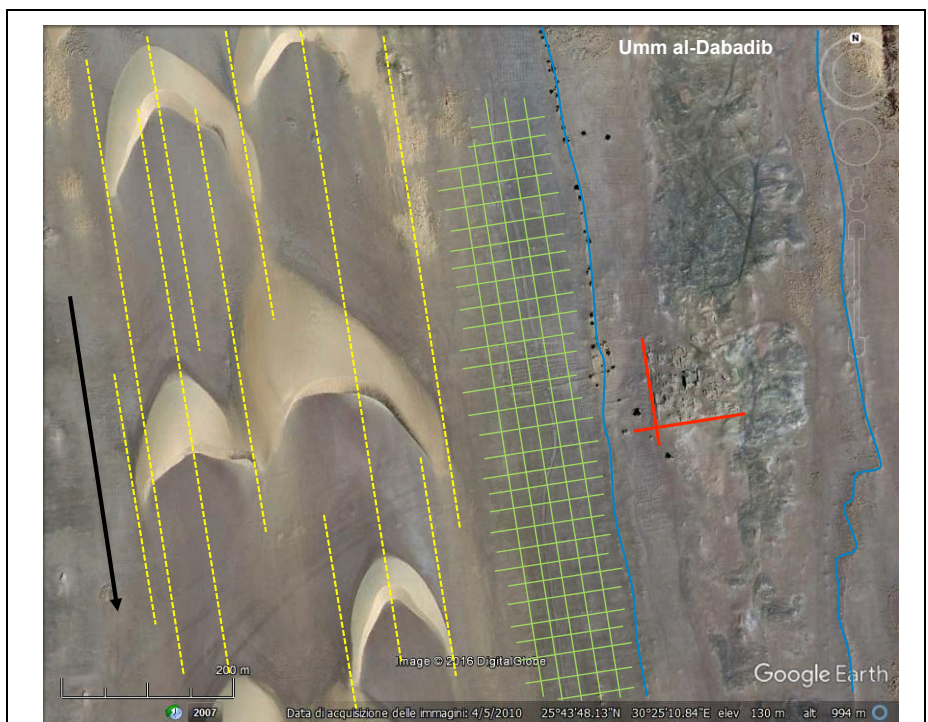

**Figure 2**: Google Earth image of the Fortified Settlement of Umm al-Dabadib and surroundings, showing the orientation of the built-up area, the *centuriatio*, the aqueducts and the sand dunes.

The entire Western Desert is actually shaped by the constant north-north-westerly wind: seen from the satellite, the entire region is crossed by an endless sequence of 'slanting', parallel lines of sand and dunes. The shape of the dunes reveals the wind pattern, that – however - can be locally affected by the presence of changes of altimetry. By looking at the way sand accumulates, one can gain a fairly good idea of the local winds, and thus compare them with the orientation of the local settlements almost at each site.

For this reason, we have reported in Table 1 the average orientation of local sand dunes. The comparison with the forts' orientations, listed from north to south, clearly shows a close correlation, with one exception: Qasr al-Gib, the northernmost fort. This building is perched on an a rock outcrop, in the midst of a rather rocky area; the orientation of this building is 3°W, and cannot be precisely matched with any topographic feature. The fort has a smaller satellite some 2 km to the south, Qasr al-Sumayra, which is also located far from any substantial accumulation of sand. The orientation of the latter is the same of the large dunes in the plain which is about 10°W.

The Fort and the Gridded Settlement of Ain al-Lebekha follow a line pointing ca 15° W; the local chain of dunes roughly follows the same direction, although the varied shape of the dunes indicate the presence of turbulences and changes of direction of the prevailing wind, due to the proximity of the escarpment and to the presence of a high massif immediately to the west of the site (figure 3 above). The large legionary fortress of al-Deir is even more clearly aligned (ca 11° W) with the dunes that proceed undisturbed far to the west in the barren plain (figure 3 below).

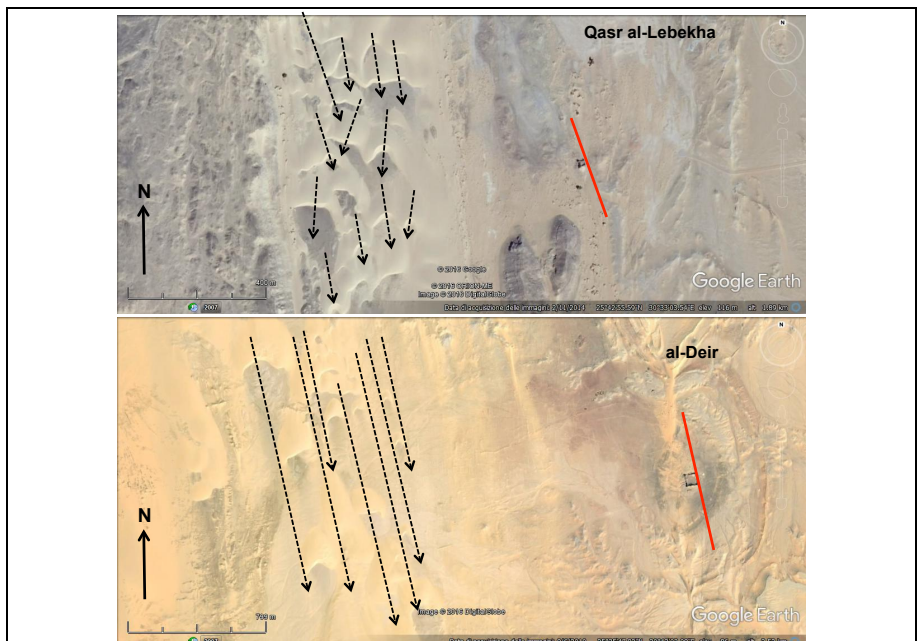

**Figure 3**: Google Earth images showing the orientation of the two Late Roman settlements of Qasr al-Lebekha and al-Deir and the sand dunes that punctuate their surroundings.

A similar situation occurs for the fort at Muhammed Tuleib, some 8.5 km to the east. No dunes are currently passing by Muhammed Tuleib, but the stripes and marks left on the surrounding landscape by the constant wind matches the orientation of the building at 10° W. This building contains the remains of an older mudbrick temple, of which, however, very little is known; the depleted remains of its eastern façade were incorporated by the outer wall of the new building, but the poor conditions of the remains do not allow a precise investigation of the relative orientation of the temple and the external building without proper archeological excavation (Ikram and Rossi 2007: 175-6).

The situation of Qasr al-Baramoudy and Qasr al-Nessima is interesting: the first is oriented 10°E and the second 7°E. They both lie at the edges of the deepest portion of the oasis' depression, near or surrounded by thick vegetation. The pattern of the sand shaped by the wind is less evident, and yet the undulating thick layer of sand that engulfs Qasr al-Baramoudy matches the slight eastward orientation of the main buildings. No sand dunes are close to Qasr al-Nessima but the orientation clearly matches that of all the cultivated fields in the area (figure 4).

The sequence of satellite images clearly shows that the orientation of the Baramoudy settlement appears to be slightly irregular and displaced a few degrees west, so that it does not match the orientation of the central building. If compared with the perfect alignment of the Fort with the Fortified Settlement at Umm al-Dabadib, this situation might indicate that the central building and settlement were not built together.

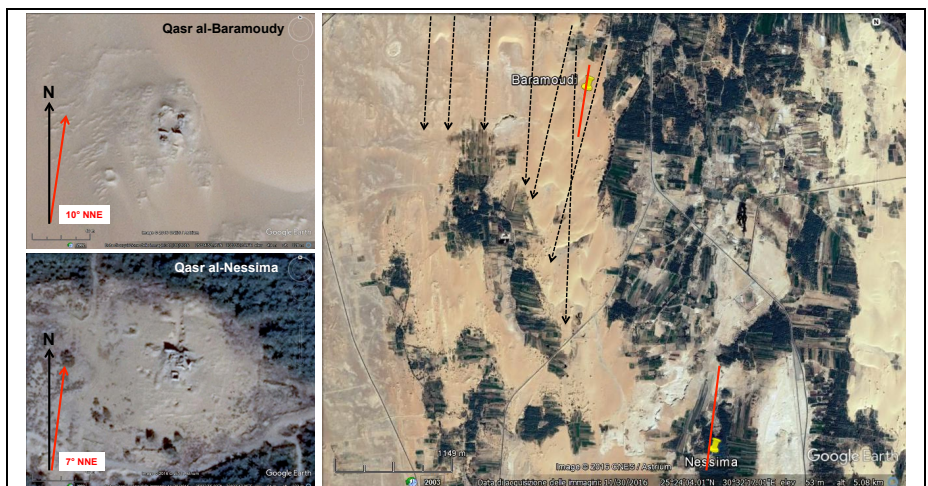

**Figure 4**: Google Earth images showing the orientation of the settlements of Qasr al-Nessima and Qasr al-Baramoudy, in central Kharga, in comparison with the local sand dunes.

In the south of the oasis, the large site of Dush represents an interesting case. Differently from the Late Roman settlements in the north, here it is clear that the built-up area is the result of a long evolution that spans several centuries, at least from the 1st to the 5th century AD. The main building, reused by the Roman army in the 4th century AD follows an orientation of ca 15°W; it contains (and is aligned to) the remains of a substantial early Roman stone temple. The nearby remains of another mudbrick temple are oriented along a similar direction, whilst the surrounding, slightly irregularly laid-out settlement follows a direction of ca 8°W (figure 5). The sand dunes that move south in the plain nearby follow directions less defined than in other areas but ranging between 8° and 15° W ca. The stratified nature of the site and the extended chronology of its occupation makes the interpretation of these remains less simple in comparison with the northern sites, built in one phase and in the same historical period. At any rate, it is interesting to note that, in this case also, the orientations of the various components of this large and complex site match that of the local wind.

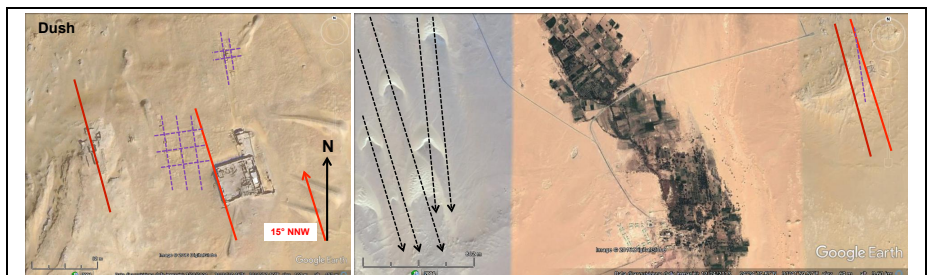

**Figure 5**: Google Earth images showing the slightly different orientations of the various elements of the built-up area of Dush, in southern Kharga, in comparison with the local sand dunes.

Finally, in central Kharga there are three sites that developed as mudbrick enclosures around pre-existing stone temples: Nadura, Qasr al-Ghweita and Qasr al-Zayyan. They are not included in the table, since it is clear that orientation of the temples – which might have depended on different factors - influenced that of the enclosures. For completeness, it is worth mentioning that their azimuth is easily measurable and essentially coincides with the one given by Belmonte and Shaltout for the temples (2006), that is ~93° E for Qasr al-Ghweita, ~108° E for Nadura and 179° E for Qasr al-Zayyan.

**Discussion: A 'Weathervane Orientation'?**

The results presented above clearly point to the conclusion that the Late Roman settlements were planned taking into account the prevailing NNW wind.

The Roman architects were well aware of the importance of the local winds when they established the orientation of their towns: Vitruvius, for instance, dedicated an entire chapter of his Book I to the winds and their dangers, stressing the necessity to avoid that they would blow across the streets (*De Architectura* I, 6). He also provides a (rather unpractical) method to establish the best orientation based on the use of a gnomon to establish the north-south direction, followed by the division of the surrounding circle into sixteenths, with the final aim to identify the eight regions of the main winds. The final recommendation is that 'the alignment of the streets and side streets ought to follow the angles between the regions of two different winds' (*De Architectura* I, 6.7, Rowland and Noble Howe 1999: 30).

In his Book XVIII, 'The Natural History of the Grain', Pliny the Elder also suggested an astronomical method to establish the four cardinal points and, as a consequence, the direction from which the winds blow locally (*Naturalis Historia* XVIII, 76. (33.): The Theory of the Winds). All these complex passages to establish first the cardinal points and then to identify the local winds can be skipped in case the wind is only one, and in case its relationship with the broader astronomical context is deemed unimportant. In this case, establishing its direction can be directly achieved with a simple weathervane.

The settlements of Late Roman Kharga appear to follow a desert-adapted version of Vitruvius' principle, that is, their orientation is based on the wind, but their aim is to exploit it, not to avoid it.

Vitruvius underlined the disadvantages of the winds ('cold winds are disagreeable, hot winds enervating, moist winds unhealthy', Book I, chapter 6.1) and recommended 'shutting out the winds from our dwellings' (6.3). This could be achieved by avoiding to lay down the streets facing the direction of the winds, and point the corners of the buildings in that direction, so that their force would be broken and dispersed (6.8). In Egypt, however, the northerly wind is generally welcome, as it provides relief from the summer heat; and even if in the desert it can be quite strong, it was perceived as a natural element to be exploited, rather than avoided. Traditionally, the main doors of the ancient Egyptian houses were located along the north side, and this tradition appears to have continued well into the Roman times, as the houses in the nearby Dakhla Oasis attest (Boozer A. L. 2016: 158).

The Late Roman forts and settlements were built not to divert the wind, but to 'filter' it.

Differently from the typical Egyptian houses, the main gate generally faced south, but this depended on a combination of reasons. Qasr al-Gib, for instance, was endowed with a single opening, the main gate facing south, which was disguised under a porch. It was a small solid building meant to act as a checkpoint along the caravan route heading north, and was not endowed with any visible opening probably to disguise from the distance its real size (Ikram and Rossi 2004: 77). The Fort at Umm al-Dabadib, instead, was probably designed to convey an aggressive appearance to travelers coming from south, and the builders felt evidently free to take full advantage of its orientation: whilst the southern front was endowed with two towers, one gate and a small number of windows, the flat northern front was pierced by at least eighteen windows, one per room. They were all small, in order to reduce the heat and the sand that would otherwise sneak inside, but must have provided a welcome draft of cool air inside the building.

Unfortunately no other northern wall survive in good conditions in the other Late Roman forts, and it is difficult to tell whether the same system was adopted elsewhere. In fact, it should be noted that the choice to orient the walls of the settlement facing the northern wind took its toll on the constructions: after sixteen centuries, nearly all the east-west walls have been eroded away by the wind-blown sand. The northern faces of forts and settlements, if they still exist, are reduced to a battered and irregular mass of mudbricks, with two exceptions: Qasr al-Gib, located on a rock outcrop and evidently less exposed to the force of the heavier sand grains blowing at ground level, and the Fort Umm al-Dabadib, where the mass of the Fortified Settlement absorbed most of the damage and protected the central Fort.

Following Vitruvius' recommendation to align the corners of the buildings towards the main wind might have protected the wall facing north from the erosion, but would have fatally diverted the evening breeze. Clearly the long-term preservation of these constructions belongs to our realm of interests, whereas the ancient builders were rightly concerned with the comfort of their constructions.

Whether or not this 'weathervane orientation' was introduced by the Romans is impossible to tell, as the pre-Roman remains (besides the temples) in Kharga are too scant and poorly documented to draw any conclusion. In any case, the same orientation appears in a clear way in the layout of the agricultural system. The seven subterranean aqueducts that provided water to the cultivations at Umm al-Dabadib were quarried along the sides of the *wadis*, to avoid both the loose filling of the lowest part and the taller rocks on the upper part. Once the water reached the surface, it was channeled towards the fields by long open-air canals, constructed by taking advantage of the local topography (lines of mounds modeled by the erosion), also shaped by the constant wind. For this reason, the aqueducts also generally appear to follow a similar alignment if compared with the Fortified Settlement and the sand dunes. The orientation of the cultivations, clearly visible from the satellite images, perfectly matches that of the Fortified Settlement, and therefore of the dunes. This layout was clearly intentional: by following the topography of the *wadis* shaped by the erosion and the plains bordered by the fields of dunes, the *centuriatio* filled the available space in the most effective way, by aligning the longest lines to the longest dimension of the terrain to be cultivated (cf. figure 2).

In conclusion, the Late Roman settlements of north Kharga were designed to exploit in the best possible way the local environmental conditions in terms of air circulation and topography. The apparent absence of astronomical references in the orientation of these settlements and, on the opposite, the strong link with the local environment suggest that this Roman large-scale investment focused on the local conditions: the optimal exploitation of the local terrain appears to have been the main concern and the focus of the builders' efforts. This represents an indirect clue on the scope of this chain of settlements, evidently meant to firmly establish the Roman control over this specific area.

Clearly, from the point of view of Archaeoastronomy, the conclusions exposed here correspond to negative results, since we have shown that astronomical orientation was *not* used. However, negative results are of course a key ingredient of the scientific method in any discipline, and to reach our conclusions we fully applied the approach of modern Archaeoastronomy, which takes into account all the aspects (natural, and man-made) of the landscape, not only of the sky.